# An efficiency analysis of Spanish airports

Adrián Nerja*

# Abstract

Privatization and commercialization of airports in recent years are drawing a different picture in the aeronautical industry. Airport benchmarking shows the accommodation and performance of airports in the evolution of the market and the new requirements that they have to face. AENA manages a wide and heterogeneous network of airports. There are 46 airports divided into three categories and with particularities due to their geographical location or the competitive environment where they are located. This paper analyzes the technical efficiency and its determinants of the 39 commercial airports of the AENA network between the years 2011-2014. To do this, two benchmarking techniques, SFA and DEA, are used, with a two-stage analysis. The average efficiency of the network is between 75-79\%. The results with the two techniques are similar with a correlation of 0.67. With regard to the commercial part of the network, AENA has a high margin for improvement because it is below the world and European average. AENA must focus on the development of the commercial area and the introduction of competition within the network to improve the technical efficiency of regional airports mainly.

Classification codes · L93, L91.

*Universitat de Valencia, Departamento de Análisis Económico, Facultat d'Economia ; adrian.nerja@uv.es

Acknowledgements: The author thanks the financial support from the Spanish Ministry of Economy, Industry and Competitiveness (BES-2014-068948). The author is grateful to professors Rafael Moner-Colonques, Jose Sempere-Monerris and Pedro Cantos Sanchez for their very helpful comments. As well as Ricardo Flores-Fillol for his constructive comments and suggestions that have substantially improved the paper.

# 1. Introduction

The airport industry has moved to a situation where every airport operates on its own. This fact has favored the entry of private capital through privatization processes. However, Spain is the only big country that maintains the joint management of a network of airports through a public company, Aeropuertos Españoles y Navigación Aérea (AENA). The decision of a joint management was justified on the grounds that complementarities would improve the competitiveness against other European airports.

However, it happens that most airports within the AENA network share a catchment area with other airports. This suggests that there may be concerns about the strategic relationship of the airports and how this may affect their efficiency. There is a debate about the closing of some airports due to their low performance, since keeping them open also affects other neighboring airports that lose passengers, which impedes the exploitation of economies of scale.

This paper analyzes the technical efficiency of AENA's airports, as well as what factors determine that efficiency. In this way, a ranking between airports can be established, in addition to observing the trend in the period analyzed. On the other hand, the endogenous and exogenous factors that explain the technical efficiency of airports are examined, since inefficiencies are not only the result of managerial decisions, but also depend on external factors.

Regarding the analysis of the technical efficiency of airports, two approaches have been employed in the literature, a non-parametric approach with Data Envelopment Analysis (DEA) and a parametric or Stochastic Frontier Analysis (SFA) approach. Liebert and Niemeier (2013) present a survey where they review more than fifty papers on benchmarking of airports in different markets using the parametric and/or the non-parametric approach. Other relevant surveys in the transport economics are, for example, De Borger et al. (2002) on public transport, Oum et al. (1999) and Cantos et al. (2012) on the rail sector, and Gonzalez and Trujillo (2009) on seaports.

Both methodologies differ in their model specification and data requirements, which can lead to different results, and that is one of the reasons why both models are used in this paper. Farrel (1957) was the first to measure the technical efficiency in a non-parametric form, which led to the development of DEA by Charnes et al. (1978). In turn, Aigner and Chu (1968) and Aigner et al. (1977) developed parametric techniques for the same purpose. Both techniques have been improved giving rise to a rich variety of possible models to be used.

The particularity of DEA is that it does not need the specification of a functional form on production, but uses linear programming to build the frontier that is determined by the most efficient airports in the sample. The main limitation of DEA is that it does not allow hypotheses to be tested. In contrast, SFA, in addition to explaining inefficiencies, incorporates stochastic random error that accounts for noise. The problem is that it is necessary to specify a functional form between inputs and outputs. There are differences between these models, although the purpose of their use is the same. In this paper, both are used to analyze possible differences in the sample and to test the robustness of the results.

Regarding the air transport market, the most relevant articles that analyze airport efficiency through the SFA are Pels et al. (2001), Pels et al. (2003) and Scotti et al. (2012). The latter use a multi-output translogarithmic function, and apply a two-stage process to analyze the competitive determinants of technical efficiency. Regarding DEA, Barros and Dieke (2008) and Barros (2008) are the closest to this paper, since they run an analysis in two stages to an output oriented function applying the techniques of Simar and Wilson (2007) for the second stage.

The literature has also focused on specifically studying the technical efficiency of Spanish airports. Martín and Roman (2001) and Tapiador et al. (2008) did an analysis applying DEA. On the other hand, Martín et al. (2009) analyzes efficiency with a cost function, while the focus of this paper is on a production function. Finally, Tovar and Martín-Cejas (2009, 2010) take an SFA with an input-oriented production function. The purpose of this paper is to analyze the technical efficiency of AENA's Spanish airports network through a two-step process using an SFA with a multi-output and output oriented production function and a DEA output oriented and variable returns to scale.

The reasons for this analysis is to analyze AENA's decisions regarding management and the privatization of the network. AENA has decided to carry out the partial privatization of the network, so that all airports operate under the same entity that makes the decisions. The literature has shown that the efficiency of airports does not depend on their ownership. On the other hand, there are other aspects that are affected by this aspect. In particular, the development of the commercial area of airports. AENA has a performance in this area below the global average and much lower than the key airports in the industry. Oum and Yu (2004) concluded that the higher the share of non-aeronautical revenues, the higher the productivity of airports. Therefore, placing the focus in this area would favor the increase of efficiency due to the complementarities in these two areas. On the other hand, it is observed how AENA increases its non-aeronautical revenues year after year. As indicated by Oum et al. (2006), private airports have to put the focus on the development of the commercial area which in turn, indirectly, improves the technical efficiency of airports. This is a key aspect to take into account in the future development of the industry and the AENA network.

The next aspect to analyze is the joint management decision. AENA has preferred to maintain the integrity of the network through the principle of solidarity between airports. However, under this decision the free competition of the airports is limited, in addition to a decision-making more adapted to the situation of each one. Tapiador et al. (2008) states an individualized management would allow a better decision-making. An individualized management would ensure competition in the market, nonexistent by how the network is formed, and the development of new market strategies.

The following section describes the current situation of the Spanish airport market. Sections 3 shows the methodology applied in the analyzes. Section 4 presents the data. In Sections 5 and 6, the estimations and the interpretation of the results of the analysis of both, airport efficiencies and their conditioning factors, are carried out. Finally, section 7 presents the conclusions and suggestions for future research.

## 2 The Spanish Airport System

Nowadays, AENA is a public company that manages the network of Spanish airports. This network consists of 46 airports and 2 heliports. In addition, AENA manages, through its international subsidiary, another 15 airports in Europe and America. AENA is organizationally attatched to Ministerio de Fomento and jointly take all management decisions such as the allocation of slots, airport charges, accounting policies, investments and fees, negotiations with airlines, etc.

In 2011 the process of privatization of the airport sector in Spain began with the creation of a mercantile society, Sociedad Mercantil AENA Aeropuertos, S.A. This company keeps exercising the functions of management of airport services, while the public entity of AENA supervises the competence in air navigation. The provision of control tower services is also liberalized in some airports, with the intention of reducing costs and increasing the competitiveness of air transport. In 2014, a partial minority privatization of 49% was made, and a 21% of which was distributed among reference partners: Corporación Financiera Alba (8%), TCI (6,5%) and Ferrovial (6,5%). The rest, 28%, went public in February 2015.

As for the data of AENA, Figure 1 shows the evolution of passenger air traffic in Spain since 2004. For the last five years of data, 2013-2017, the number of passengers has increased by 33%, to reach almost 250 million.

*Figure 1: AENA passengers between 2004-2017*

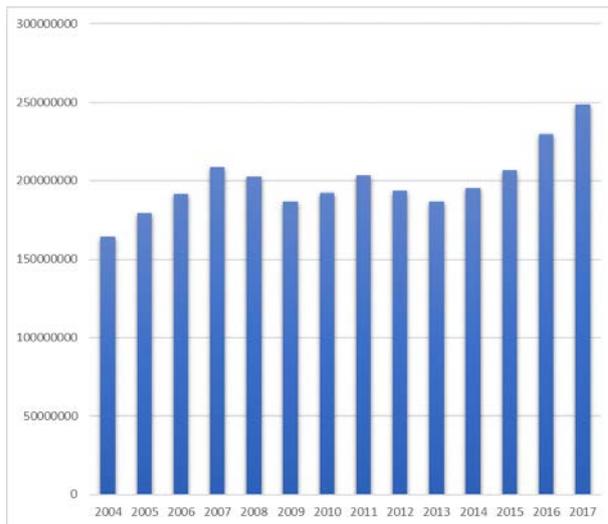

The decision of the Spanish government was to maintain an integrated and centralized airport system. This decision is different from that adopted by many other countries such as France, Italy, Germany and the UK in Europe, and Canada and the United States in America. These countries decided to establish a decentralized management of airports. Spain opted for centralization to avoid the possible negative effects of competition between the airports of the network, and by the position of AENA in Latin America.

Another main reason for maintaining joint and centralized management is due to the single or common fund. This allows AENA to take advantage of economies of scale when accessing the credit for investments. However, the most important thing is that there is a policy of solidarity and redistribution, where financial resources are transferred from profitable to non-profitable airports. This fact can raise incentive problems. Airports with losses have no incentives to reduce them, while those with profits also have no incentive to generate more profitability if they cannot appropriate it. It should be pointed out that the surplus far exceeds the amount of deficit airports.

This measure on cross-subsidies is usually used to ensure the connectivity of remote areas where there is no alternative, also known as public services obligation routes (PSOs). But that is not the case of Spain, especially in the peninsula. This has led to an excess of small underutilized airports, affecting public financial resources (European Commission, European Court of Auditors, 2014). This decision, therefore, affects the efficiency of airports and the network, and has raised a question about airport competition in Spain.

To analyze the competition in the Spanish airport network, the literature distinguishes three types of competition:[1]

***Competition for catchment area or spatial competition.*** 26 airports in the network have at least one airport less than 130 km away, as can be seen in Figure 2. This indicates that there is a strong competitive factor to attract the potential passengers within the catchment area.

---

[1] Extracted from Santaló and Socorro (2015).

*Figure 2: Network of airports managed by AENA*

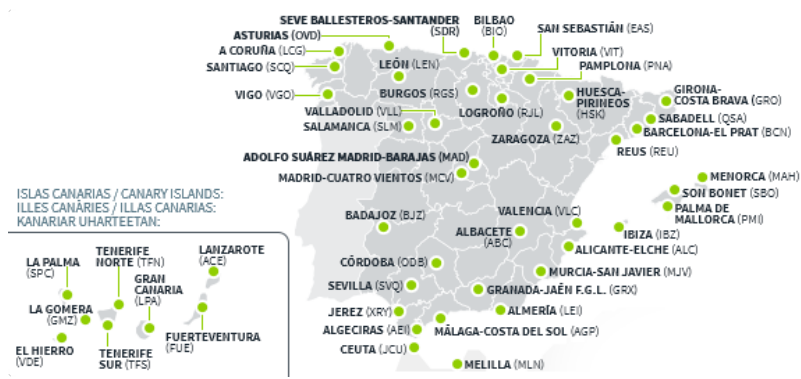

**Competition for tourist destinations**. Spain has 14 tourist airports. These airports face two types of competition. Competition within the Spanish market, and competition with other international destinations.

**Competition for connection traffic.** This is the competence of hub airports. In Spain between Adolfo Suárez Madrid-Barajas and Barcelona-El Prat.

Within the efficiency analysis, privatization does not guarantee that better results are achieved. However, competition may encourage higher levels of efficiency. An appropriate way to achieve competition would be through an individualized management of airports. In this way, each airport would look after its interests taking the most appropriate measures to capture traffic and to generate a positive impact in the territory where it is located.

Airport competition brings benefits for all agents in the economy. The most directly affected are the airlines, since they receive a better service and reduce the costs of the derived services. This fact has a direct impact on airfares, as long as the airlines pass on to their customers the reduction in costs. In this case, the demand would increase, producing a positive externality in the commercial part of the airport.

It is our purpose to analyze the efficiency of the airports in the AENA network. As indicated by Martín et al. (2009), "there are always different stakeholders, viz., regional planners, regulators or investors who need information on the cost structure and efficiency of airports". That is to say, the analysis of the efficiency has policy and managerial implications and will become more relevant with potential changes in the air industry.

# 3 Methodology

SFA and DEA are two widely used methods in the literature and both allow the inclusion of multiple inputs and output to calculate production frontiers and measure efficiency with respect to the constructed frontier. The efficiency analysis carried out in this paper is derived from two sources: the estimation of the efficiency values and the analysis of determinants of efficiency. Each of them is presented below following the two approaches previously named.

## 3.1 Estimation of parametric efficiency scores

A stochastic distance function econometric model is used to obtain the technical efficiencies of airports. There are some issues when applying DEA that SFA overcomes, such as the sensitivity to the number of inputs and outputs used relative to the number of observations, the presence of outliers affecting the efficiency levels of the rest of airports, and the lack of dynamism and justification behind the efficiency scores, (Zarraga, 2017). The function used is the Translog Multioutput Distance Function that reflects how many inputs are used to generate the outputs. Then, the more resources the airport has, the more production can be obtained.

The data set does not include monetary variables, but physical inputs and outputs with the aim of measuring technical efficiency. In this case, an output oriented stochastic distance function seems to be more appropriate than an input oriented. This is because, at least in the short run, many inputs in airport operation are indivisible.

In this framework, the airports' production possibility set is defined as $P(x)$ - i.e., the output vector $y \in R_+^M$ that can be obtained using the input vector $x \in R_+^M$, that is: $P(x) = y \in R_+^M: \{x \ can \ produce \ y\}$. By assuming that the technology, $P(x)$, satisfies the axioms listed in Fare et al. (1994), the Shephard (1970) output-oriented distance function is introduced:

$$D_0(x,y) = \min\left\{\theta: \left(\frac{y}{\theta}\right) \epsilon P(x)\right\},$$

where $\theta \leq 1$. The distance function is non-decreasing, positively linearly homogeneous, and convex in $y$, and decreasing in $x$, Lovell et al. (1994). $D_0(x,y) = 1$ means that $y$ is located on the outer boundary of the production possibility set - i.e., $D_0(x,y) = 1 \ if \ y \epsilon IsoqP(x) = \{y: y\epsilon P(x), \omega y \notin P(x), \omega > 1\}$. If instead $D_0(x,y) =< 1$, $y$ is located below the frontier; in this case, the distance represents the gap between the observed output and the maximum feasible output.

Following Coelli and Perelman (2000), the translog distance function is given by:

$$\ln(D_{0it}/y_{Mit}) = \alpha_0 + \sum_{m=1}^{M-1} a_m \ln y_{mit}^* + \frac{1}{2} \sum_{m=1}^{M-1} \sum_{n=1}^{M-1} a_{mn} \ln y_{mit}^* \ln y_{nit}^* + \sum_{k=1}^{K} \beta_k \ln x_{kit}$$
$$+ \frac{1}{2} \sum_{k=1}^{K} \sum_{l=1}^{K} \beta_{kl} \ln x_{kit} \ln x_{lit} + \sum_{k=1}^{K} \sum_{n=1}^{M-1} \gamma_{km} \ln x_{kit} \ln y_{mit}^*,$$

(1)

where $i$ denotes the $ith$ airport in the sample. $M$ is the number of outputs denoted by $y$, and $K$ is the number of inputs denoted by $x$. $D_{0it}$ is the output distance from the frontier of firm $i$ in period $t$ and $y_{mit}^* = y_{mit}/y_{Mit}$. Symmetry is imposed as:

$a_{mn} = a_{nm}; m, n = 1,2, \ldots, M, and \ \beta_{kl} = \beta_{lk}; k, l = 1,2, \ldots, K.$

The required restrictions for homogeneity of degree +1 in outputs are:

$$\sum_{m=1}^{M} a_m = 1,$$

$$\sum_{n=1}^{M} a_{mn} = 0, m = 1, 2, \ldots, M,$$

$$\sum_{m=1}^{M} \gamma_{km} = 0, k = 1, 2, \ldots, K.$$

(2)

Then, Equation (1) can be written as $\ln(D_{0it}/y_{Mit}) = TL(x_{it}, y_{mit}/y_{Mit}, \alpha, \beta, \gamma)$, where $TL$ stands for the translog function. Hence:

$$-\ln(y_{Mit}) = TL(x_{it}, y_{mit}/y_{Mit}, \alpha, \beta, \gamma) - \ln(D_{0it})$$

*(3)*

where, $\ln(D_{0it})$ is non-observable and can be interpreted as an error term in a regression. If it is replaced with $(v_{it}, u_{it})$, the typical SFA composed error term is achieved: $v_{it}$ are random variables that are assumed to be *iid* as $N(0, \sigma_u^2)$ and independent of the $u_{it}$; the latter are non-negative random variables distributed as $N(m_{it}, \sigma_u^2)$. The inefficiency scores are given by $u_{it}$, whereas $v_{it}$ represent the random shocks. Hence, the translog output oriented stochastic function for estimation can be written as:

$$-\ln(y_{Mit}) = \alpha_0 + \sum_{m=1}^{M-1} a_m \ln y^*_{mit} + \frac{1}{2} \sum_{m=1}^{M-1} \sum_{n=1}^{M-1} a_{mn} \ln y^*_{mit} \ln y^*_{nit} + \sum_{k=1}^{K} \beta_k \ln x_{kit} + \frac{1}{2} \sum_{k=1}^{K} \sum_{l=1}^{K} \beta_{kl} \ln x_{kit} \ln x_{lit} + \sum_{k=1}^{K} \sum_{n=1}^{M-1} \gamma_{km} \ln x_{kit} \ln y^*_{mit} + v_{it} - u_{it}$$

*(4)*

To investigate the determinants of inefficiency, a single-stage estimation procedure following Battese and Coelli (1992) is applied, where the technical inefficiency effect, $u_{it}$, in Eq. (3) can be specified as:

$$u_{it} = \{\exp[-\eta(t - T)]\} u_{it}$$

*(5)*

This model is such that the non-negative inefficiencies, $u_{it}$, decrease, remain constant or increase as $t$ increases, if $\eta > 0, \eta = 0$ or $\eta < 0$, respectively. The parameter $\eta$ is an unknown scalar, whereas $T$ represents the time periods. The case in which $\eta$ is positive is likely to be appropriate when firms tend to improve their level of technical efficiency over time. This equation characterizes the improving learning curve over time.

The technical efficiency of airport $i$ at period $t$ is defined as follows:

$$TE_{it} = e^{-u_{it}}$$

*(6)*

where $0 \leq TE_{it} \leq 1$.

### 3.1.1 Tobit analysis of determinants of efficiency

Regression analysis is often applied in a second stage in which the efficiency estimate is regressed against a set of potential variables. Then, a Tobit regression model is used since the dependent variable has a left limit of 0 and an upper limit of 1. Thus, the Tobit model is represented in the following equation:

$$TE_{it} = \alpha + \beta_k x_{itk} + \varepsilon_{it}$$

*(7)*

where $TE_{it}$ represents the efficiency estimated for airport $i$ at time $t$, $x_i$ is used to interpret the dependent variables, $\alpha$ is a constant term, $\beta_k$ indicates the parameter entries for each explanatory variable, and $\varepsilon_{it}$ is a random error vector.

## 3.2 Estimation of non-parametric efficiency scores

Charnes et al. (1981) describe the DEA methodology as "a mathematical programming model applied to observational data that provides a new way of obtaining empirical estimates of external relations; such as the production functions and/or efficient production possibility surfaces that are a cornerstone of modem economics."

Originally, DEA was designed as a non-parametric method of frontier estimation to evaluate the relative efficiency of decision-making units (DMUs), which use multiple inputs to produce multiple outputs, without a clear identification of the relation between them. DEA assumes neither a specific functional form for the production function nor the inefficiency distribution. There are two basic DEA models; variable returns to scale (VRS), and constant return to scale (CRS). DMUs that do not lie on the frontier are inefficient and the measurement of the degree of inefficiency is determined by the selection of the model.

There must be a good understanding over the data set used. It is especially important to have some idea about the hypothetical returns to scale that exist in the industry, because this is going to determine the envelopment surface-constant return to scale CRS (Charnes et al., 1978) or variable return to scale VRS (Banker et al., 1984) of the model. For the purpose of this analysis, VRS is chosen because there are airports of different sizes and types, that is, with different scales.

Furthermore, there are two main orientations: input and output. As explained for the SFA model, an output-oriented model is used because of the indivisibility of airport inputs in the short run.

The standard procedure in the non-parametric approach is followed. For each period $t$, there is a set of $N$ airports ($i = 1, \ldots, N$). Each airport produces a vector of $y_{im}$ outputs ($m = 1, \ldots, M$) using $x_{ik}$ inputs ($k = 1, \ldots, K$). The measurement of each airport's efficiency is obtained through the comparison with a linear combination of the total number of airports included in the analysis. Formally, the output-oriented DEA efficiency for airport $i$ is calculated through the following linear programming problem:

$$Max \; \theta_j^{VRS}$$
$$s.t. \sum_i \lambda_i y_{im} \geq \theta_j y_{jm} \; \forall m,$$
$$\sum_i \lambda_i x_{ik} \geq y_{jk} \; \forall k,$$
$$\sum_i \lambda_i = 1$$
$$\lambda_i \geq 0$$

*(8)*

The solution of this problem gives $N$ optimal values for $\theta^{VRS}$, that is, the airports efficiencies when using VRS and an output oriented analysis. The efficiencies are expressed in such a way that $0 \leq \theta^{VRS} \leq 1$.[2] Those airports with $\theta^{VRS} < 1$ are considered inefficient, while they are efficient when meet $\theta^{VRS} = 1$. The parameter λ shows the weights assigned to each airport to perform the analysis.

## *3.2.1 Simar-Wilson analysis of determinants of efficienty*

For this case, the approach of Simar and Wilson (2007) has been used. A common practice in the DEA literature has been the use of the Tobit-estimator until demonstrated the inadequacy of such approach. Therefore, those papers that have used a two-step analysis are invalid according to Simar and Wilson (2007), who showed a data generation process (DGP) that is consistent when using a second stage. This approach is based on a truncated-regression with a bootstrap process. Then, the econometric model is given by:

$$\theta_{it}^{VRS} = \alpha + Z_{it}\delta_i + \varepsilon_{it}, i = 1, \ldots, n,$$

(9)

where,

$$\varepsilon_{it} \sim N(0, \sigma_\varepsilon^2), such\ that\ \varepsilon_{it} \geq 1 - \alpha - Z_{it}\delta_i, i = 1, \ldots, n$$

(10)

---

[2] The efficiency measure considered as the technical efficiency is calculated as the inverse of the maximum proportional output that can be obtained for the indicated inputs, $\frac{1}{\theta_j^{VRS}}$.

In equation (9), $\theta_{it}^{VRS}\epsilon(0,1]$ is the efficiency calculated with the DEA technique, $\alpha$ is the constant term, $\varepsilon_{it}$ is the statistical noise and $Z_{it}$ is a vector that contains the variables that try to explain the efficiency. Since $\theta_{it}^{VRS}$ is bounded by unity, $\varepsilon_{it} \geq 1 - \alpha - Z_{it}\delta_i$, and is assumed that the distribution is truncated normal with zero mean.

## 4. Data

Spain has a total of 51 airports, 46 of which are managed by AENA and the other are privately owned or inoperative. Figure 2 shows the map of the AENA airport network in Spain, while Table 1 shows a classification of these airports according to their typology.

From table 1 it can be seen that AENA has two heliports and several air military bases that it manages together with the Spanish Armed Forces. Some of these airports do not have civil traffic. For the analysis of this paper, there is a total of 38 airports in the AENA network, where heliports and general aviation airports are excluded from the analysis. That is, the list is reduced to hub, touristic and regional airports.[3] The data included in the analysis are collected from the public information of AENA.

*Table 1: Types of airports managed by AENA*

| Type of Airports | Number of airports |
|---|---|
| Hub: Adolfo Suárez Madrid-Barajas and Barcelona-El Prat. | 2 |
| Touristic: Alicante-Elche, Almería, Fuerteventura, Girona-Costa Brava, Gran Canaria, Ibiza, La Palma, Lanzarote, Málaga-Costa del Sol, Menorca, Palma Mallorca, Reus, Tenerife Sur and Valencia. | 14 |
| Regional: A Coruña, Albacete, Asturias, Badajoz, Bilbao, Burgos-Villafría, El Hierro, FGL Granada-Jaén, Jerez, La Gomera, León, Logroño-Agoncillo, Melilla, Murcia-San Javier, Pamplona, Salamanca, San Sebastián, Santander, Santiago, Sevilla, Tenerife Norte, Valladolid, Vigo, Vitoria and Zaragoza. | 25 |

---

[3] Albacete and Melilla are excluded because of their very low activity. Vitoria is also excluded because it is specialized in cargo instead of passengers.

| | |
|---|---|
| Heliport: Algeciras and Ceuta. | 2 |
| General aviation: Córdoba, Huesca-Pirineos, Madrid-Cuatro Vientos, Sabadell and Son-Bone. | 5 |

For the first stage where airport efficiency is analyzed, a balanced panel with a total of 152 observations is used, since the data collected is from 38 airports in the 2011-2014 period. This period has been chosen because of data availability. A multi-input/multi-output distance function output oriented with three outputs and three inputs is used. Table 2 shows a summary of the descriptive statistics of these six variables. These variables are broadly used in the literature as Liebert and Niemeier (2013) present in their survey. The financial variables, NAR and EMP, are deflated and expressed in 2011 euros.

*Table 2: Descriptive Statistics of Output (O) and Input (I) Variables*

| | **Mean** | **Std. Dev.** | **Min** | **Max** |
|---|---|---|---|---|
| ATM (O) | 47669,78 | 76213,18 | 1201 | 429390 |
| SIZE (O) | 73,66409 | 40,7989 | 2,290864 | 153,504 |
| NAR (O) *mill.* | 17,92023 | 37,44989 | 0,01 | 207,0323 |
| EMP (I) *mill.* | 9,548649 | 12,30422 | 0,460333 | 79,58212 |
| RUNW (I) *meters* | 3534,414 | 2511,259 | 1250 | 15450 |
| TERM (I) $m^2$ | 90938,5 | 190723,8 | 2326 | 955305 |

Three outputs are considered. The annual number of aircraft movements (ATM), the average size of aircraft (SIZE), and the non-aeronautical revenue (NAR). Regarding the ATM, Figure 3 shows its evolution since 2004 and reflects the negative impact caused by the financial crisis in Spain in 2007. Nowadays, ATM has exceeded 2004 values, but has not yet achieved full recovery, though a similar trend will allow to close the gap shortly.

*Figure 3: Air Transport Movements between 2004-2017*

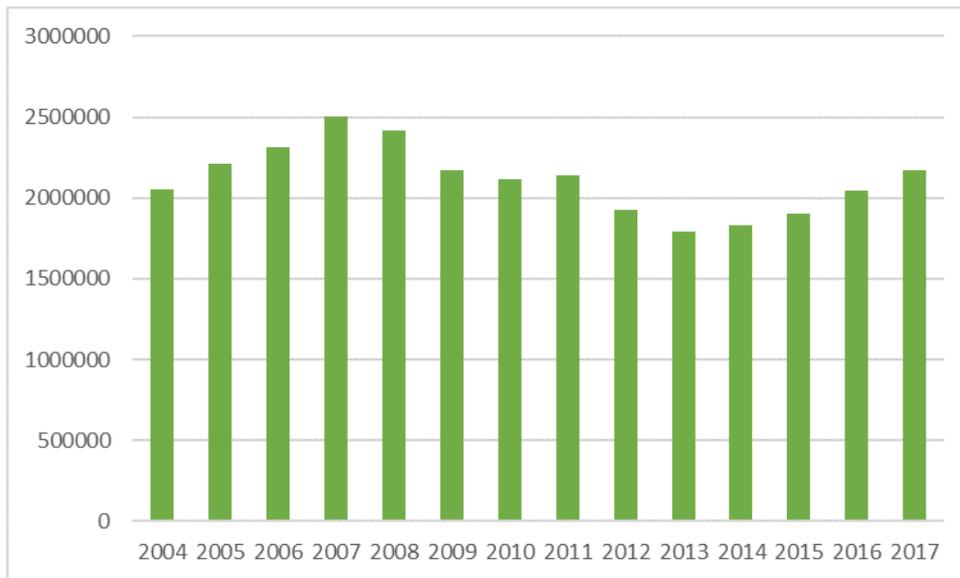

The average size of aircraft is defined as the ratio between yearly passenger volume and ATM, such as *SIZE=PAX/ATM*[4]. This output is taken into account to measure the relative efficiency of each airport due to the fact that in the sample there are airports of different categories. In hubs, or connecting airports, larger aircraft usually allow greater movement of people and, therefore, a more efficient behavior. This contrasts with the type of existing aircraft in regional airports that are smaller and with a much lower capacity.

The non-aeronautical revenue, NAR, is also analyzed. Increasingly, airports are relying on this type of income to make improvements and investments. It is therefore important to control for this element as it is an indicator of how efficiently an airport uses its facilities and the behavior of passengers. AENA, compared to other airports, is not well positioned with respect to the commercial development of airports, which indicates that there is a great potential to be exploited.

With regard to inputs, the deflated costs of each airport in employees (EMP), the length of runways in meters (RUNW), and the area in square meters of the airport terminal (TERM) are taken into account. Regarding employment, we do not have data on the number of employees per airport, however, since it is a single company that manages all workers and is under a public management, an approximation is made using the employee costs of each airport. It is understood that all workers, no matter which airport they are, belonging to the same company, have equal salaries.

As a measure to analyze the operational capacity of the airport, the runway length is used. 31 of the 38 airports analyzed have only one runway. The Adolfo Suárez Madrid-Barajas airport has four runways, the most numerous, followed by Barcelona-El Prat, which has three. Finally, the area of the terminal, TERM, is used. This variable directly influences the commercial development of airports. The growth of terminals and runways is not linear, but occurs in a discrete way. Therefore, decisions about investing in these variables is important because it affects the efficiency of airports in the short and long term.

---

[4] This specification of the variable SIZE is used by Oum et al. (2006) and Tovar and Martín-Cejas (2009, 2010) among others.

*Table 3: Pearson Correlation of Output (O) and Input (I) Variables*

|  | ATM (O) | SIZE (O) | NAR (O) | EMP (I) | RUNW (I) | TERM (I) |
|---|---|---|---|---|---|---|
| ATM (O) | 1,0000 |  |  |  |  |  |
| SIZE (O) | 0,5192 | 1,0000 |  |  |  |  |
| NAR (O) | 0,9706 | 0,5078 | 1,0000 |  |  |  |
| EMP (I) | 0,9594 | 0,6015 | 0,9461 | 1,0000 |  |  |
| RUNW (I) | 0,8937 | 0,4221 | 0,8910 | 0,8719 | 1,0000 |  |
| TERM (I) | 0,9470 | 0,4770 | 0,9724 | 0,9312 | 0,8821 | 1,0000 |

Table 3 shows the Pearson correlation coefficients. It is observed how the inputs are correlated showing a congruence between the operative and functional part of the airports. It is also observed how the outputs ATM and NAR are related to each other. This correlation is to be expected since a greater number of passengers leads to the commercial revenues of the airport growing. On the other hand, there is a correlation between outputs and inputs showing that airports try to make an optimum use of their facilities.

# 5. Efficiency of airports

This section analyzes the results regarding the efficiency scores of airports. In this first stage, the technical efficiencies are obtained by applying the techniques described above, SFA and DEA.

## 5.1 SFA results

The following multi-output stochastic distance function is estimated in order to obtain the efficiency values:

$$-\ln(ATM_{it}) = TL(SIZE_{it}/ATM_{it}, NAR_{it}/ATM_{it}, EMP_{it}, RUNW_{it}, TERM_{it}, \alpha, \beta, \gamma) + v_{it} - u_{it}$$

(11)

where $ATM_{it}$ is the normalizing output (i.e., $SIZE_{it}$ an $NAR_{it}$d are expressed in $ATM_{it}$ terms), $\alpha$ is a vector of coefficients for $SIZE_{it}/ATM_{it}$ and $NAR_{it}/ATM_{it}$. $\beta$ is a vector of coefficients regarding inputs, and $\gamma$ is a vector of coefficients related to output-input interactions.

Table 4 presents the results. Of the first order coefficients, the only one that is significant is non-aeronautical revenue *(NAR ')*, which demonstrates the strong relationship between the number of flights and commercial revenues obtained by an airport. It is logical to think that the more people attending an airport, the non-aeronautical revenues increase by the complementarity between both demands. Concerning second-order coefficients, the size of the aircraft *(SIZE ')*, non-aeronautical revenues *(NAR')*, and the runway length *(RUNW)* are statistically significant. Furthermore, some interaction effects are statistically significant.

Table 4: SFA estimates

| Parameter | Estimate | Est. Error |
|---|---|---|
| *Constant* | 17,573 | 13,090 |
| SIZE' | 1,000 | 1,065 |
| NAR' | 2.800*** | 1,043 |
| EMP | 1,161 | 0,849 |
| RUNW | 1,887 | 2,157 |
| TERM | -0,544 | 0,960 |
| SIZE'2 | -0.165*** | 0,056 |
| NAR'2 | 0.092* | 0,050 |
| SIZE' x NAR' | 0,027 | 0,039 |
| EMP2 | -0,158 | 0,099 |
| RUNW2 | -0.601*** | 0,207 |
| TERM2 | -0,076 | 0,052 |
| EMP x RUNW | 0,130 | 0,116 |
| EMP x TERM | -0.100** | 0,052 |
| RUNW x TERM | 0,051 | 0,129 |
| EMP x SIZE' | -0,067 | 0,042 |
| EMP x NAR' | 0.129* | 0,069 |
| RUNW x SIZE' | -0,111 | 0,145 |
| RUNW x NAR' | -0,171 | 0,158 |
| TERM x SIZE' | -0.150*** | 0,045 |
| TERM x NAR' | -0,043 | 0,064 |
| Log-likelihood | 135,461 | |
| N obs | 152 | |

Note *, **, *** denote significance at 10%, 5% and 1% respectively; and ' denotes normalized output (SIZE/ATM and NAR/ATM).

The technical efficiency scores of airports per year are gathered in Table 5. The average is around 0.8, which matches with Tovar and Martín-Cejas (2009, 2010). Even so, there are significant differences between airports, since there are airports that almost reach the unit with very little inefficiency, while there are cases that are only 30% efficient. The trend has been decreasing in the years of the sample. On average, a 3.24% efficiency was lost. The airport that has seen its efficiency reduced the most has been Logroño. As can be seen in Figure 2, Logroño shares a catchment area with five airports. And is the airport, together with Bilbao, with most surrounding airports. On the other hand, those that have suffered the least losses have been the most efficient airports.

*Table 5: Airport's Technical SFA Efficiency Scores*

| Airport | 2011 | 2012 | 2013 | 2014 | Mean | % Var |
|---|---|---|---|---|---|---|
| A Coruña | 0,870 | 0,865 | 0,859 | 0,853 | 0,862 | -1,90 |
| Alicante | 0,962 | 0,960 | 0,959 | 0,957 | 0,959 | -0,529 |
| Almería | 0,677 | 0,665 | 0,653 | 0,641 | 0,659 | -5,24 |
| Asturias | 0,853 | 0,848 | 0,841 | 0,835 | 0,844 | -2,16 |
| Badajoz | 0,718 | 0,708 | 0,697 | 0,686 | 0,702 | -4,46 |
| Barcelona/El Prat | 0,845 | 0,839 | 0,832 | 0,825 | 0,835 | -2,29 |
| Bilbao | 0,885 | 0,880 | 0,875 | 0,870 | 0,877 | -1,67 |
| Burgos | 0,615 | 0,602 | 0,589 | 0,575 | 0,596 | -6,48 |
| El Hierro | 0,964 | 0,963 | 0,961 | 0,960 | 0,962 | -0,49 |
| Fuerteventura | 0,961 | 0,960 | 0,958 | 0,956 | 0,959 | -0,53 |
| Girona | 0,925 | 0,922 | 0,918 | 0,915 | 0,920 | -1,06 |
| Gran Canaria | 0,846 | 0,840 | 0,834 | 0,827 | 0,837 | -2,27 |
| Granada/Jaén/FGL | 0,845 | 0,839 | 0,832 | 0,826 | 0,836 | -2,29 |
| Ibiza | 0,955 | 0,953 | 0,951 | 0,949 | 0,952 | -0,629 |
| Jerez | 0,664 | 0,652 | 0,640 | 0,628 | 0,646 | -5,49 |
| La Gomera | 0,427 | 0,411 | 0,395 | 0,379 | 0,403 | -11,10 |
| La Palma | 0,644 | 0,631 | 0,619 | 0,606 | 0,625 | -5,90 |
| Lanzarote | 0,975 | 0,974 | 0,973 | 0,972 | 0,973 | -0,341 |
| León | 0,418 | 0,402 | 0,386 | 0,370 | 0,394 | -11,36 |
| Logroño | 0,340 | 0,324 | 0,308 | 0,293 | 0,316 | -13,85 |
| Madrid /Barajas | 0,933 | 0,930 | 0,927 | 0,924 | 0,929 | -0,944 |
| Málaga/Costa del Sol | 0,889 | 0,885 | 0,880 | 0,875 | 0,882 | -1,60 |
| Menorca | 0,941 | 0,939 | 0,936 | 0,934 | 0,938 | -0,82 |
| Murcia/San Javier | 0,976 | 0,975 | 0,973 | 0,972 | 0,97 | -0,33 |
| Palma de Mallorca | 0,980 | 0,979 | 0,978 | 0,977 | 0,978 | -0,27 |
| Pamplona | 0,454 | 0,438 | 0,422 | 0,407 | 0,430 | -10,34 |
| Reus | 0,832 | 0,826 | 0,819 | 0,812 | 0,822 | -2,49 |
| Salamanca | 0,587 | 0,574 | 0,560 | 0,546 | 0,567 | -7,08 |
| San Sebastián | 0,846 | 0,840 | 0,833 | 0,827 | 0,836 | -2,28 |
| Santander | 0,961 | 0,960 | 0,958 | 0,956 | 0,959 | -0,53 |
| Santiago | 0,924 | 0,920 | 0,917 | 0,914 | 0,919 | -1,08 |
| Sevilla | 0,804 | 0,796 | 0,788 | 0,780 | 0,792 | -2,96 |
| Tenerife Norte | 0,831 | 0,824 | 0,817 | 0,810 | 0,820 | -2,52 |
| Tenerife Sur | 0,956 | 0,954 | 0,952 | 0,950 | 0,953 | -0,61 |
| Valencia | 0,751 | 0,742 | 0,732 | 0,722 | 0,737 | -3,87 |
| Valladolid | 0,912 | 0,908 | 0,904 | 0,900 | 0,906 | -1,25 |
| Vigo | 0,785 | 0,776 | 0,768 | 0,759 | 0,772 | -3,29 |
| Zaragoza | 0,953 | 0,951 | 0,949 | 0,947 | 0,950 | -0,65 |
| **Average** | **0,808** | **0,801** | **0,795** | **0,788** | **0,798** | **-3,24** |

Figure 4 shows a ranking of airports with respect to their TE in 2014. The five most efficient airports in 2014 with more than 0.95 of average TE are Palma de Mallorca, Murcia/San Javier, Lanzarote, El Hierro, and Alicante. On the other hand, the three most inefficient are La Gomera, León, and Logroño. Airports are inefficient by definition, but you can find cases with very significant inefficiencies. Eliminating some exception, these airports are located in the interior of the peninsula. They are regional airports in areas of low tourist interest, with low population densities, and that often share an area of influence with other nearby airports. On the other hand, it can be seen how the first nine airports are located on the coast. These airports are located in areas of high tourist interest. It seems that an important element is the location and here in Spain the tourist attraction of sun and beach prevails.

*Figure 4: Ranking of airports by Technical Efficiency in 2014*

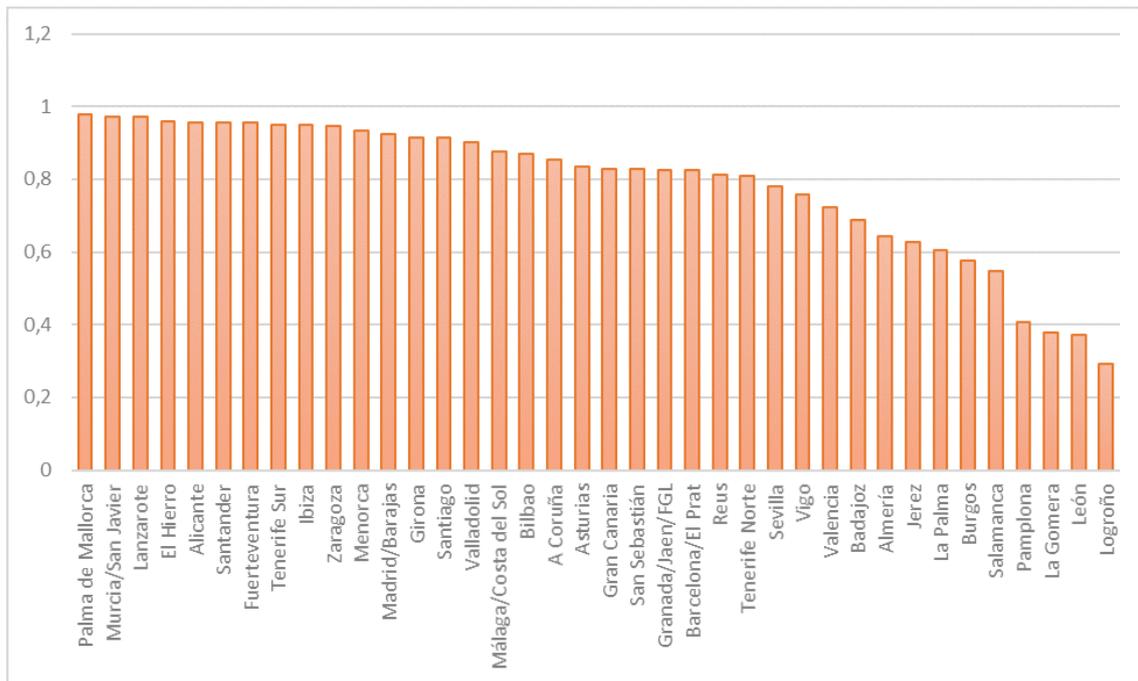

## 5.2 DEA results

When obtaining the values of efficiency using the non-parametric DEA technique, the three inputs and the three outputs previously exposed are used without altering. In this case, the imposition of any functional form is not required. Table 6 shows the technical efficiencies obtained under this technique.

DEA efficiencies can be calculated in different ways. In this study, efficiencies are estimated through an output-oriented approach, assuming that airports aim to maximize the profits they obtain from their activities. In addition, variable returns to scale between inputs and outputs are assumed following the model presented by Banker et al. (1984).

*Table 6: Airport's Technical DEA Efficiency Scores*

| Airport | 2011 | 2012 | 2013 | 2014 | Mean | %Var |
|---|---|---|---|---|---|---|
| A Coruña | 0,827 | 0,755 | 0,809 | 0,866 | 0,814 | 4,72 |
| Alicante | 0,977 | 0,988 | 1 | 1 | 0,991 | 2,30 |
| Almería | 0,564 | 0,486 | 0,695 | 0,676 | 0,605 | 19,86 |
| Asturias | 0,803 | 0,877 | 0,884 | 0,805 | 0,843 | 0,229 |
| Badajoz | 0,795 | 1 | 0,823 | 1 | 0,904 | 25,64 |
| Barcelona/El Prat | 1 | 0,987 | 0,989 | 1 | 0,994 | 0 |
| Bilbao | 1 | 0,837 | 1 | 0,998 | 0,958 | -0,17 |
| Burgos | 1 | 0,640 | 0,622 | 0,626 | 0,722 | -37,33 |
| El Hierro | 1 | 0,600 | 0,952 | 1 | 0,888 | 0 |
| Fuerteventura | 0,922 | 0,851 | 0,975 | 0,994 | 0,936 | 7,78 |
| Girona | 0,880 | 0,832 | 0,964 | 0,861 | 0,884 | -2,17 |
| Gran Canaria | 1 | 0,920 | 0,934 | 0,913 | 0,942 | -8,61 |
| Granada/Jaén/FGL | 0,660 | 0,538 | 0,607 | 0,600 | 0,601 | -9,12 |
| Ibiza | 1 | 0,918 | 0,984 | 1 | 0,975 | 0 |
| Jerez | 0,993 | 0,922 | 1 | 0,911 | 0,956 | -8,26 |
| La Gomera | 0,570 | 0,316 | 0,484 | 0,551 | 0,480 | -3,37 |
| La Palma | 0,619 | 0,569 | 0,635 | 0,646 | 0,617 | 4,30 |
| Lanzarote | 1 | 0,961 | 1 | 1 | 0,990 | 0 |
| León | 0,420 | 0,326 | 0,404 | 0,419 | 0,393 | -0,20 |
| Logroño | 0,343 | 0,286 | 0,321 | 0,367 | 0,329 | 6,85 |
| Madrid /Barajas | 1 | 1 | 0,961 | 1 | 0,990 | 0 |
| Málaga/Costa del Sol | 1 | 0,856 | 0,860 | 0,876 | 0,898 | -12,37 |
| Menorca | 0,693 | 0,684 | 0,799 | 0,813 | 0,747 | 17,29 |
| Murcia/San Javier | 1 | 0,977 | 1 | 1 | 0,994 | 0 |
| Palma de Mallorca | 1 | 0,992 | 1 | 1 | 0,998 | 0 |
| Pamplona | 0,403 | 0,299 | 0,430 | 0,401 | 0,383 | -0,36 |
| Reus | 0,765 | 0,621 | 0,721 | 0,698 | 0,701 | -8,78 |
| Salamanca | 1 | 0,722 | 1 | 0,825 | 0,886 | -17,46 |
| San Sebastián | 1 | 1 | 0,969 | 1 | 0,992 | 0 |
| Santander | 0,760 | 0,678 | 0,812 | 0,736 | 0,747 | -3,13 |
| Santiago | 0,810 | 0,764 | 0,848 | 0,831 | 0,813 | 2,57 |
| Sevilla | 0,790 | 0,681 | 0,730 | 0,739 | 0,735 | -6,55 |
| Tenerife Norte | 0,979 | 0,798 | 1 | 0,912 | 0,922 | -6,86 |
| Tenerife Sur | | 0,974 | 0,979 | 1 | 1 | 0,988 | 2,65 |
| Valencia | | 1 | 0,837 | 0,890 | 0,897 | 0,906 | -10,24 |
| Valladolid | | 0,732 | 0,592 | 0,721 | 0,699 | 0,686 | -4,47 |
| Vigo | | 0,673 | 0,612 | 0,620 | 0,613 | 0,629 | -8,94 |
| Zaragoza | | 0,587 | 0,474 | 0,532 | 0,504 | 0,524 | -14,05 |
| **Average** | | **0,830** | **0,741** | **0,815** | **0,810** | **0,799** | **-1,79** |

DEA efficiency scores are between zero and 1. Airports with DEA scores equal to 1 are efficient. Any airport with a score less than one is relatively inefficient, that is, an airport with a score of 0.9 is only 90% efficient as the best performing airport. Efficiencies scores are not absolute, but are relative because they depend on the rest of airports. Having said that, significant differences can be observed between Spanish airports as in previous research (Martín and Roman, 2001, and Tapiador et al., 2008).

Despite these differences, the average is almost 80%. However, many airports have a high level of inefficiency, as shown in Figure 5. In addition, the variation in efficiency, although negative, has been low. Although it should be noted that some airports decrease their efficiency, while others increase it. It would be expected that smaller airports would be the ones that increase the efficiency, since their margin of improvement and growth is greater, but this does not happen.

By comparing the results with those obtained with SFA, it is observed that they are similar with a correlation of 0.66. In addition, on average it can be seen that the value of efficiency is very similar (0.798 in SFA against 0.799 in DEA) and the variation has been also negative.

*Figure 5: Ranking of airports by DEA efficiency averages*

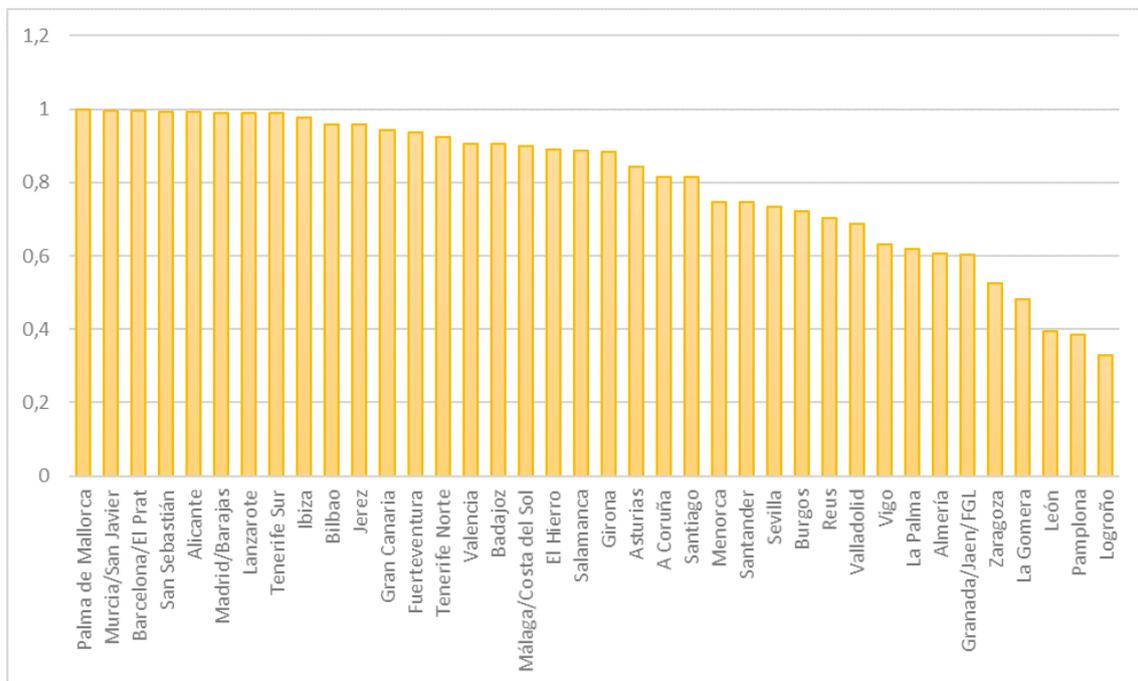

As can be seen in Figure 5, the five most efficient airports on average and with more than 0.99 are Palma de Mallorca, Murcia/San Javier, Barcelona/El Prat, San Sebastian, and Alicante; the most inefficient are León, Pamplona, and Logroño. As noted, there is some difference, although many airports coincide comparing these results with the ones of the parametric distance function.

# 6. Determinants of efficiency

In the second stage, the determinants that explain the technical efficiency of airports are analyzed. Financial variables, SUB and EBITDA, are deflated in prices of 2011. Therefore, the dependent variables in this analysis are the efficiency estimates calculated in the first stage. Table 7 summarizes the descriptive statistics of the variables used in the second-stage analysis.

*Table 7: Descriptive Statistics of the second stage*

| Variable | Obs | Mean | Std Dev | Min | Max |
|---|---|---|---|---|---|
| $TE_{SFA}$ | 152 | 0,7984425 | 0,1815546 | 0,2932976 | 0,9800863 |
| $\theta_{DEA}$ | 152 | 0,7995202 | 0,2028277 | 0,286638 | 1 |
| ISLE | 152 | 0,2894737 | 0,4550173 | 0 | 1 |
| HUB | 152 | 0,0526316 | 0,2240351 | 0 | 1 |
| TOUR | 152 | 0,3684211 | 0,483971 | 0 | 1 |
| CONG *units* | 152 | 1,526316 | 1,395107 | 0 | 5 |
| LCC | 152 | 0,5855263 | 0,4942595 | 0 | 1 |
| SUB *mill.* | 152 | 1,031532 | 1,406445 | -0,2623907 | 8,619001 |
| EBITDA *mill.* | 152 | 35,07294 | 94,51639 | -7,74 | 555,0049 |
| HSR | 152 | 0,1973684 | 0,3993285 | 0 | 1 |
| HH | 152 | 3290,86 | 2478,323 | 483,772 | 9701,45 |

The explanatory variables used are ISLE, which is a dummy variable that indicates whether the airport is located on an island. In total, there are eleven airports located on islands and it is intended to analyze if there is a difference with the airports in the hinterland, since the catchment area in islands is limited by their extension.

Following the type of airports in Table 1, dummies of each category are introduced, where regional airports are used as a variable of control. Therefore, the dummies that are introduced are HUB and TOUR, for hub and tourist airports. In Figure 2, the relation of airports and their geographic disposition is shown. Then, a variable is introduced that measures the conglomeration of airports, CONG. Conglomeration is measured as the number of airports that share a catchment area within a 150 km radius. As can be seen, every airport has at least one close airport, whereas Bilbao and Logroño have five airports less than 150 km away.

Because low-cost carriers have influenced the aeronautical industry, a dummie, LCC, is introduced. This variable indicates when the airline that moves the most passengers at the airport is a low cost carrier. Many airports have survived thanks to LCC. As observed in Table 7, in more than half of airports it is an LCC that attracts the most traffic.

To measure the bargaining power of the airport against airlines, the Herfindahl-Hirschman index, HH, is used. This index measures the concentration of the airlines within each airport by passenger volume, and ranges from 0 to 10,000, being 10,000 the value that marks the monopoly of the market, that is, a single airline operating in one airport. It is expected that a higher level of concentration will give airlines greater bargaining power. For example, La Gomera is the most concentrated airport with almost absolute monopoly, 9701 from 10000. However, AENA manages all airports in the network, so the bargaining power is centralized and not individualized. Airports cannot negotiate terms or conditions unilaterally with airlines.

This fact reduces the chance of airlines to increase bargaining power in an isolated airport. It should be noted that La Gomera flights is under the PSO regulation.

In order to control for some financial variables, the subsidies that airports receive, SUB, are introduced. Since airports are public, they usually have grants from different agencies. It is observed that, on average, airports receive one million euros a year of subsidies. The EBITDA is also introduced, which shows an indicator of the operating profitability of the business without taking into account any financial aspects. Of the 152 data, 67 are negative. In addition, 12 airports have a negative EBITDA in the entire period analyzed, compared to 17 that have a positive EBITDA. However, there is an average of more than 35 million euros of exploitation. This is due to the fact that those with a positive EBITDA generate, on average, more than 66 million euros, while those that generate losses are only slightly less than 3 million euros. This fact reinforces the theory that cross-subsidization does not affect the network because the airports that make losses, lose very little, and those that win, earn a lot. If net effects are taken into account, it is still profitable to maintain the current system.

Due to the emergence of multimodal competition, and that Spain is, after China, the country with the largest High Speed Rail (HSR) network, we introduce a variable that is activated when the city airport has a HSR station. In Spain there is no airport that has an integrated HSR station, but there are eight cities that have a HSR station. To be sure, the competition between the HSR and the regional flights has had a real impact, and it is reasonable to analyze if this fact affects the technical efficiency of airports.

Finally, three temporal dummies (T11, T12, T13) are introduced to analyze the evolution of technical efficiency and if there is some event that requires a more exhaustive analysis. The control dummy corresponds to the year 2014; the last period of the panel.

## 6.1 Discussion of results

In the second stage, the determinants of technical efficiency are analyzed. Once obtained the technical efficiencies under each model used in stage one, the data is analyzed, although with a different treatment in each case. In addition, the function that is analyzed is the following:

$$Eff = f(ISLE, HUB, TOUR, COMP, NARPAX, LCC, HH, SUB, EBITDA, HSR, T11, T12, T13)$$

(12)

To perform the analysis of the *TE* obtained under the parametric distance function, a Tobit regression with panel data is carried out, where the dependent variable is the technical efficiency obtained in the first stage through SFA.

On the other hand, to perform the analysis of the DEA estimates, the technique of Simar and Wilson (2007) is applied. The technical efficiencies under DEA are not parametric. In addition, they are censored since there are usually several DMUs with value one, and there is a problem of correlation between the estimated values. They propose a double bootstrap procedure that improves statistical efficiency in the second-stage regression.

Table 8 shows the results of the estimates for each model specified above:

*Table 8: Second Stage Regression*

|  | **Tobit (SFA)** | **Simar Wilson (DEA)** |
|---|---|---|
| *Constant* | 0.844*** | 0.165*** |
| ISLE | 0.025*** | -0.163*** |
| HUB | 0.054* | -1.547* |
| TOUR | 0.033*** | -0,071 |
| CONG | -0.029*** | -0.077*** |
| LCC | 0,001 | 0.107*** |
| SUB | 0,001 | 0,024 |
| EBITDA | 0.000*** | 0.008*** |
| HSR | 0.010** | -0.186*** |
| HH | -0.006*** | -0.089*** |
| T11 | 0.022*** | 0,037 |
| T12 | 0.015*** | -0,027 |
| T13 | 0.008*** | 0,034 |

Note *, **, *** denote significance at 10%, 5% and 1% respectively.

The variables that have the same relationship in both models for significance and sign are the competition between airports *(COMP)*, the pre-tax benefits *(EBITDA)* and the concentration of the airlines at the airport *(HH)*. The rest of variables, except the subsidies *(SUB)*, are significant in one of the two models applied. Public subsidies do not affect the efficiency of airports because it is an extended practice among all airports, so everyone is in the same situation. Looking at Table 6, the standard deviation is 1.40, while the average is over 1 million euros per year. Except for some airports that have received significant amounts in some periods, the rest receive a systematic subsidy around the average.

Attending to the AENA network intracompetition, Figure 2 shows the airports in the network and how many of them share a catchment area with neighboring airports. That is what the variable *COMP* measures, the number of adjoining airports within a 150 km radius. Competition is expected to improve efficiency because airfares are lowered and more passengers are attracted; instead, the opposite is observed. Note that, in this case, there is no competition because it is the same company who manages the whole network of airports, therefore that competitive component is not transferred to the prices or to the relationship with airlines. The result is that there is a loss of efficiency. Airports instead of accessing new passengers, they have to share the existing demand in the same catchment area.

AENA decided not to individualize the management of the airports so as not to disadvantage the airports with less activity, the regional ones, and that could continue operating within the network. This decision monopolizes the principle of solidarity between airports, but keeps alive the debate about the closure of some airports. In the peninsula, the geographical location of airports is relevant. Those located on the coast enjoy the competitive advantage of tourism. On the other hand, the regional airports that are in the interior do not have this advantage, so the competition factor to have airports in their same catchment area affects them negatively. Tapiador et al. (2008) supports the hypothesis that an individualized management would help airports compete under their particular characteristics. So competition between neighboring airports may also encourage the development of new market strategies, reaching the situation in which the relationship of this variable with the efficiency of airports is positive.

The geographical characteristics explain why tourist airports are more efficient than regional airports. This is to be expected since the tourist airports are quite specialized and attract a specific type of passengers. In addition, regional airports have the problem that they attract less demand and this explains their lower technical efficiency. Both of them are usually point-to-point airports that can optimize their resources much better than hub airports. Hub airports are more inefficient than regional ones according to the non-parametric analysis. This is due to the fact that hub airports have to satisfy a more heterogeneous demand. Besides, investments are scattered, which makes the optimization of resources more complicated.

Airports located on islands also have a geographical limitation. They are less efficient due to the fact that they lack of the potential to reach more customers. Investments at airports are discrete, so there is always a tendency to overinvest so that demand is adjusted step by step to supply. In addition, it is the case that in almost all the islands there is at least one airport, which limits even more the power of attraction between neighboring islands.

Another important aspect that affects the performance of airports is the structure of the air market. From a theoretical point of view, Basso and Zhang (2007) began to consider the influence of the downstream airline market structure on the performance and decision-making of airports. The different changes in the industry have caused the relationship between airports and airlines to be influenced (see D'Alfonso and Nastasi, 2014). Therefore, it is interesting to analyze how the air structure within the airport affects its efficiency. The Herfindahl-Hirschman Index shows the concentration of airlines within each airport. With this variable it is observed how airlines power affects the technical efficiency of each airport. The result is that less concentrated airports are more efficient, that is, the concentration of the airlines causes inefficiency. AENA manages airports with equal policies depending on their type, therefore, there are few chances for airlines to bargain and influence decisions. The interaction of the airlines within the airport can favor this result, since in a concentrated market, entry barriers to other airlines can be imposed, fostering greater inefficiency or a worse use of airport resources.

Another aspect that influences the structure of the air market is the type of airports. In Spain there are only two hub airports, the rest are regional and tourist airports where most of the flights are point-to-point. Under this model, low-cost companies have expanded. Then, the low-cost model is quite widespread in the AENA network, also worldwide. Therefore, it was an element to control to analyze its relevance. It is observed that the airports in which the leading airline is a low-cost, are more efficient according to the non-parametric approach. This may be due to the fast turn-around and the sharing of different aeronautical services with other airlines.

The next element to analyze, which is very timely nowadays, is intermodal competition, *HSR*. The parametric analysis indicates that it does not affect the efficiency, this is due to the fact that HSR stations are not integrated in airports; contrary to what happens in other countries. On the other hand, the non-parametric analysis produces a negative and significant result. This indicates that those airports whose cities have an HSR station are more inefficient this meaning that the substitution effect is higher than the complementary effect.

The value of *EBITDA*, although significant, is practically zero. Even so, the result was to be expected, that is, more efficient airports are expected to be the most profitable. Hub and tourist airports generate positive EBITDA, in contrast to regional, whose EBITDA is close to zero or even slightly negative.

Finally, the evolution of technical efficiency is analyzed where it is observed that it is decreasing over time. Table 6 also shows this trend, where during the period analyzed efficiency is systematically lost in all airports. It is also observed how this does not happen with the non-parametric analysis, since in that case this pattern was not found and therefore there is no significance in this second stage as the results show. Perhaps this tendency is the product of the chosen type of analysis.

The variable *NARPAX* measures the non-aeronautical revenue airports make per passenger. This measure is relevant because it indicates the behavior of passengers inside the airport. In the Tobit analysis it is very significant and has a negative effect, indicating that airports with higher non-aeronautical revenues are less efficient. The airports most focused on their commercial part have larger facilities that do not directly affect technical efficiency. That is, they could work with smaller facilities, although they would eliminate the commercial part of the business.

# 7. Conclusion and future research

Air transport is the gateway to the world of millions of passengers whose number is increasing every year. The aeronautical industry has undergone an evolution fulfilling the demand of society to have a more connected world. It is an essential key piece in the development of communications and the global economy.

It should also be noted that it is a relatively young market that is expanding and constantly evolving. It is therefore important to analyze its structure and performance, due to the impact that is generated in many sectors such as tourism, business development, technology, freight transport, etc. In addition to that, there are various interested groups affected by the performance of the industry such as passengers, public institutions and investors, among others.

For Spain, it is an essential sector. A country with approximately 46 million inhabitants that moves around 200 million air passengers per year. In recent years, important decisions have been taken in the Spanish air sector. Given the need for expansion and access to private capital to undertake investments, decisions have been made regarding privatization and the management model. Finally, it was decided to bet on a joint management model and access the partial privatization of the company that manages the Spanish airport network, AENA.

Given these decisions, it is important to analyze the behavior of airports and check if the reasons for these decisions correspond to actual data. To this end, an analysis of the technical efficiency of airports and their determinants is carried out. Two techniques are used to offer a more conclusive result, an SFA and an DEA. It also includes an analysis of the commercial area of airports, due to the significant relevance that this part of the business is acquiring.

The main limitation of this analysis has been the availability of data, that only allows to analyze a period from 2011 to 2014. To solve this problem, two different techniques have been used that provide robustness to the analysis.

Several conclusions can be drawn, although perhaps the most relevant is that the existence of joint management of airports affects negatively the technical efficiency. The majority of airports share catchment area with other airports. However, this competitive pressure that, in theory, positively affects passengers, does not exist. Therefore, the growth of demand due to the competitive effect does not occur, affecting technical efficiency negatively. This invigorates the debate about whether some regional airports should remain open or not.

The separation of airports into three types does not fit the reality. Each airport has its particularities and an individualized management would allow the adoption of more precise measures that favor a better performance of airports. Maintaining a global network ensures the continuity of some airports, but in turn limits their growth and high levels of inefficiency are obtained. Perhaps, a formula that allows greater power in individual decision-making while maintaining the joint network would be a possible solution to assess. This would favor competition within the network and the specialization of some airports, which would benefit passengers and entities.

As future research, due to the particularity of the AENA network, the analysis can be extended by type of airport, the existing competition inside and outside the network, the geographical situation, the state of the economy by regions, etc.